# Direct observation of corner states in second-order topological photonic crystal slabs

Xiao-Dong Chen[†], Wei-Min Deng[†], Fu-Long Shi[†], Fu-Li Zhao, Min Chen, and Jian-Wen Dong[*]

*School of Physics & State Key Laboratory of Optoelectronic Materials and Technologies, Sun Yat-sen University, Guangzhou 510275, China.*

[†]These authors contributed equally to this work

[*]Corresponding author: dongjwen@mail.sysu.edu.cn

**Recently, higher-order topological phases that do not obey the usual bulk-edge correspondence principle have been introduced in electronic insulators and brought into classical systems, featuring with in-gap corner/hinge states. In this letter, using near-field scanning measurements, we show the direct observation of corner states in second-order topological photonic crystal slabs consisting of periodic dielectric rods on a perfect electric conductor. Based on the generalized two-dimensional Su-Schrieffer-Heeger model, we show that the emergence of corner states roots in the nonzero edge dipolar polarization instead of the nonzero bulk quadrupole polarization. We demonstrate the topological transition of two-dimensional Zak phases of PC slabs by tuning intra-cell distances between two neighboring rods. We also directly observe in-gap one-dimensional edge states and zero-dimensional corner states in the microwave regime. Our work presents that the PC slab is a powerful platform to directly observe topological states, and paves the way to study higher-order photonic topological insulators.**

Topological insulators (TIs) host robust edge states predicted by the bulk-edge correspondence principle: A $d$-dimensional ($d$D) TI with $d$D insulating bulk states supports ($d$-1)D conducting edge states [1-3]. This bulk-edge correspondence principle explains some topological properties, such as the soliton formation in polyacetylene based on the 1D Su-Schrieffer-Heeger (SSH) model [4] and protected gapless edge states in 2D TIs [5, 6]. Recently, the concept of higher-order TIs that do not obey the usual form of the bulk-edge correspondence principle have been introduced [7-24]. For example, a second-order 2D TI has gapped 1D edge states but gapless 0D corner states. After the proposal in electronic insulators, higher-order topological phases have been realized in classical systems without the limitation imposed by the Fermi level. So far, 0D corner states have been observed in second-order insulators with the quantized bulk quadrupole polarization [15-17] or with the quantized edge dipolar polarization [12, 20, 21]. However, the latter kind of corner states have not been experimentally observed in photonic systems.

Photonic crystals (PCs) are periodic optical structures in which many fancy photonic phenomena such as negative refraction [25], cloaking effect [26], and broadband angular selectivity [27] were observed. With tunable geometric structures and controllable band dispersions, PCs provide a good platform to emulate topological behaviors. For example, researchers have successfully observed the photonic analog of quantum Hall, spin Hall, and valley Hall effect in PCs [28-37]. However, most reported 2D topological PCs require a metallic cover in experiment to prevent the radiation of electromagnetic waves into free space. This metallic cover not only complicates the experimental setup, but also hampers the direct mapping of electromagnetic fields. In view of demands, a compact platform without the metallic cover is needed to realize the direct observation of topological states.

Here, we report the direct observation of corner states in second-order topological PC slabs consisting of periodic dielectric rods on a perfect electric conductor (PEC). Our designed PC slab is free of the metallic cover, enabling the direct observation of topological states. Moreover, this structure can not only double the effective height of rods but also possess a full band gap, as compared with free-standing PC slabs. Based on the generalized 2D SSH model, we show that the emergence of corner states roots in the nonzero edge dipolar polarization. We demonstrate the topological transition of 2D Zak phases by expanding or shrinking four dielectric rods in the unit cell. By using near-field scanning measurements, in-gap 1D edge states and 0D corner states are directly visualized.

We consider the free-standing square-lattice PC slab whose unit cell consists of four close-packed dielectric rods with the permittivity of $\varepsilon = 9.5$ in the air [Fig. 1(a)]. The in-plane lattice constant is $a = 25$ mm, the height and diameter of rods are $h = 50$ mm and $d = 5$ mm. Figure 1(c) shows its bulk band structures in which TM-like and TE-like bands are marked in blue and red. The TM-like band gap ranges from 4.55 to 5.34 GHz with a 16.0% gap width. However, the large height-diameter ratio ($h/d = 10$) increases the difficulty of sample fabrication. In addition, there is no full band gap if both TM-like and TE-like modes are considered. These two disadvantages can be overcome by considering PC slab with rods on a PEC [Fig. 1(b)]. First, the effective height of rods is double with the introduced PEC boundary. To see this, we consider rods with $h = 25$ mm on a PEC. Figure 1(d) shows its bulk band structures which are exactly the same as TM-like bands in Fig. 1(c). It confirms that the effective height of rods is double because $h = 25$ mm in Fig. 1(b) while $h = 50$ mm in Fig. 1(a). Second, the PEC boundary requires that electric fields are perpendicular to the boundary, i.e., $E_x = E_y = 0$. Hence, TE-like modes are filtered out and TM-like modes are kept [inset of Fig. 1(d)]. Hence, a full band gap is found (shaded in blue). To prove the theoretical proposal, we construct the experimental sample in which periodic ceramic rods with $h$ = 25 mm are put on a metallic plate [Fig. 1(e)]. This metallic plate behaves as a PEC in the microwave regime. A source antenna is inserted through the drilled hole in the metallic substrate to excite eigen-states of PC slabs. The signal is measured by a probe antenna along the $z$ direction, and measured $E_z$ fields are collected by the vector network analyzer (Agilent E5071C). By performing the Fourier transform on scanned $E_z$ fields, we obtain the measured bulk band structures which are in good agreement with the calculated band structures [Fig. 1(f)].

The above PC slab is the photonic realization of the 2D SSH model in which the intra-cell distances between two neighboring rods are given by $d_x$ and $d_y$ [Fig. 2(a)]. By considering the first bulk band, the topology of PC slab is given by the 2D Zak phase $\mathbf{Z} = (Z_x, Z_y)$:

$$Z_j = \int dk_x dk_y \, \text{Tr}[\hat{A}_j(k_x, k_y)] , \tag{1}$$

where $j = x$ or $y$, $\hat{A}_j(k_x, k_y) = i \langle u(\mathbf{k}) | \partial_{k_j} | u(\mathbf{k}) \rangle$ with $|u(\mathbf{k})\rangle$ is the periodic Bloch function. Note that, the 2D Zak phase is related to the 2D bulk polarization via $Z_j = 2\pi P_j$ [24, 38]. For PC slabs with the mirror symmetry along the $j$ direction, $Z_j$ is quantized to 0 or $\pi$ and $Z_j$ is related to the symmetry of Bloch modes at the zone center and boundary [39]. As the first bulk state at the Γ point is mirror symmetric, $Z_x$ is 0 ($\pi$) when the bulk state at the $X$ point is mirror symmetric (mirror anti-symmetric). The case of $Z_y$ is similar by considering the bulk state at the $Y$ point. Take the PC slab with $d_x$ = 5 mm and $d_y$ = 5 mm as an example [left panel of Fig. 2(c)]. As the bulk state at the $X$ ($Y$) point is mirror symmetric [left insets of Fig. 2(d)], $Z_x$ ($Z_y$) is 0 (0). Then this PC slab (named as PCS1) is characterized by $\mathbf{Z} = (0, 0)$. We can tune $d_x$ or $d_y$ to achieve topological phase transitions. Figure 2(d) plots the frequency spectra of two lowest bulk states at the $X$ and $Y$ points by increasing $d_y$ from 5 to 20 mm while keeping $d_x$ = 5 mm. The mirror symmetric and anti-symmetric states are labelled in red and blue, respectively. The lowest bulk state at the $X$ point keeps mirror-symmetric as there is no mode exchange.

On the contrary, two bulk states at the *Y* point move closer, touch, and separate with the increasing of $d_y$. After the mode exchange at $d_y = 12.5$ mm, PC slab is characterized by $Z_y = \pi$ as the lowest bulk state at the *Y* point becomes mirror anti-symmetric. One representative PC slab with $d_x = 5$ mm and $d_y = 20$ mm is shown in the middle panel of Fig. 2(c) and named as PCS2. Similarly, PC slabs with $Z_x = \pi$ can be obtained when $d_x > 12.5$ mm. Because both $Z_x$ and $Z_y$ are either 0 or $\pi$, PC slabs are classified into four different topological phases [Fig. 2(b)]. Particularly, we can obtain PC slabs with nonzero Zak phase along both two directions, e.g., PC slab with $d_x = 20$ mm and $d_y = 20$ mm [right panel of Fig. 2(c) and named as PCS3]. To see the phase transition from PCS1 to PCS3, we consider PC slabs with $d_x = d_y$. Figure 2(e) plots the frequency spectra of two lowest bulk states at the *X* and *Y* points. After the mode exchange at both two *k*-points, PC slabs are characterized by $\mathbf{Z} = (\pi, \pi)$ as two bulk states become mirror anti-symmetric [right insets of Fig. 2(e)].

The topological distinction between PC slabs with different $Z_j$ guarantees the existence of edge states. As $Z_y$ is 0 for PCS1 while it is $\pi$ for PCS2 (or PCS3), there are edge states along the *x* direction for the boundary between PCS1 and PCS2 (or PCS3). To verify this prediction, we first construct the boundary between PCS1 and PCS2 [Fig. 3(a)]. We put the source antenna at the left and scan $E_z$ fields [see Fig. S1]. By performing the Fourier transform of measured $E_z$ fields, the dispersion of edge states is shown by the bright color in Fig. 3(b). The measured edge dispersion agrees well with the simulated dispersion marked by the green line. Gapped edge states are confirmed as the dispersion does not cover the whole bulk band gap. In addition, with measured $E_z$ fields at the frequencies of 4.39 GHz (at $k_x = 0$) and 5.18 GHz (at $k_x = \pi/a$), we can retrieve the eigen-fields of edge states at the zone center and boundary [Fig. 3(c)]. Both edge states are mirror symmetric, indicating the zero edge dipolar polarization. To have a nonzero edge dipolar polarization, we consider the boundary between PCS1 and PCS3 [Fig. 3(d)]. We also scan $E_z$ fields [see Fig. S3] and obtain the measured edge dispersion which is in good agreement with the calculated dispersion [Fig. 3(e)]. Gapped edge states are found but their dispersion curve is different to that in Fig. 3(b), indicating that they have different features. To see this, we retrieve the eigen-fields of edge states at $k_x = 0$ at the frequency of 4.41 GHz and at $k_x = \pi/a$ at the frequency of 4.84 GHz [Fig. 3(f)]. Edge state at the zone center is mirror symmetric while that at the zone boundary is mirror anti-symmetric. It confirms the nonzero edge dipolar polarization.

The nonzero edge dipolar polarization between PCS1 and PCS3 indicates the existence of corner states. Hence, we construct the experimental sample for observing corner states [Fig. 4(a)]. It consists of PCS3 at the center while PCS1 at the background. In experiment, we put a source antenna near the top-right corner. Figure 4(b) shows measured $E_z$ fields at one example frequency of 4.01 GHz [see more $E_z$ fields in Fig. S3]. To identify bulk, edge, and corner states, we apply three binary filters [lower panel of Fig. 4(b)] to separate responses of different eigen-states. Figure 4(c) shows resulting spectra in which high values indicate the existence of eigen-states and zero (or low) values mean the presence of band gaps. Within the bulk and edge gaps, a resonance of the corner spectrum is found near 5.2 GHz. It corresponds to the corner state whose existence has been predicted according to the nonzero edge dipolar polarization [Fig. 4(d)]. The zoom-in $E_z$ fields clearly demonstrate that fields of the corner state are concentrated around the top-right single rod [left panel of Fig. 4(e)]. The field concentration can be also achieved by considering the frequency within the bulk, edge, and corner gaps, e.g., localized state at f = 4.98 GHz [right panel of Fig. 4(e)]. But these two field concentrations are different because corner states are immune to the source position while localized states are dependent on the source position. Lastly, we carry out a comparative study of the sample without corner states [Fig. 4(f)]. Within the bulk and edge gaps, there is no resonance of corner spectra. It proves the absence of corner states due to the zero edge dipolar polarization between PCS1 and PCS2. More detailed discussion including the presence (absence) of edge states at the top (right) boundary is given in Fig. S4.

In conclusion, we design and demonstrate PC slabs with periodic dielectric rods on a PEC. Based on the 2D SSH model, we illustrate the topological phase map of 2D Zak phases and the topological phase transition. Inherent from the nonzero Zak phase of bulk states, we directly observe corner states induced by the nonzero edge dipolar polarization. This is the smoking-gun feature of second-order topological PC slabs in which 0D corner states are predicted by the nontrivial Zak phase of 2D bulk states. Our results demonstrate that second-order photonic TIs can guide the light flow and trap the light. Corner states may have the potential application in enhancing light-matter interaction, which is desirable for many nanophotonic devices.

*Note added*: During the submission of this manuscript, we found the simultaneous and independent discovery of corner states and higher order topological states in 2D PCs [40] and photonics waveguides [41].

*Acknowledgments*: This work was supported by National Natural Science Foundation of China (Grant Nos. 11704422, 61775243, 11522437, 61471401, and 11761161002), Natural Science Foundation of Guangdong Province (Grant No. 2018B030308005), and Science and Technology Program of Guangzhou (Grant No. 201804020029).

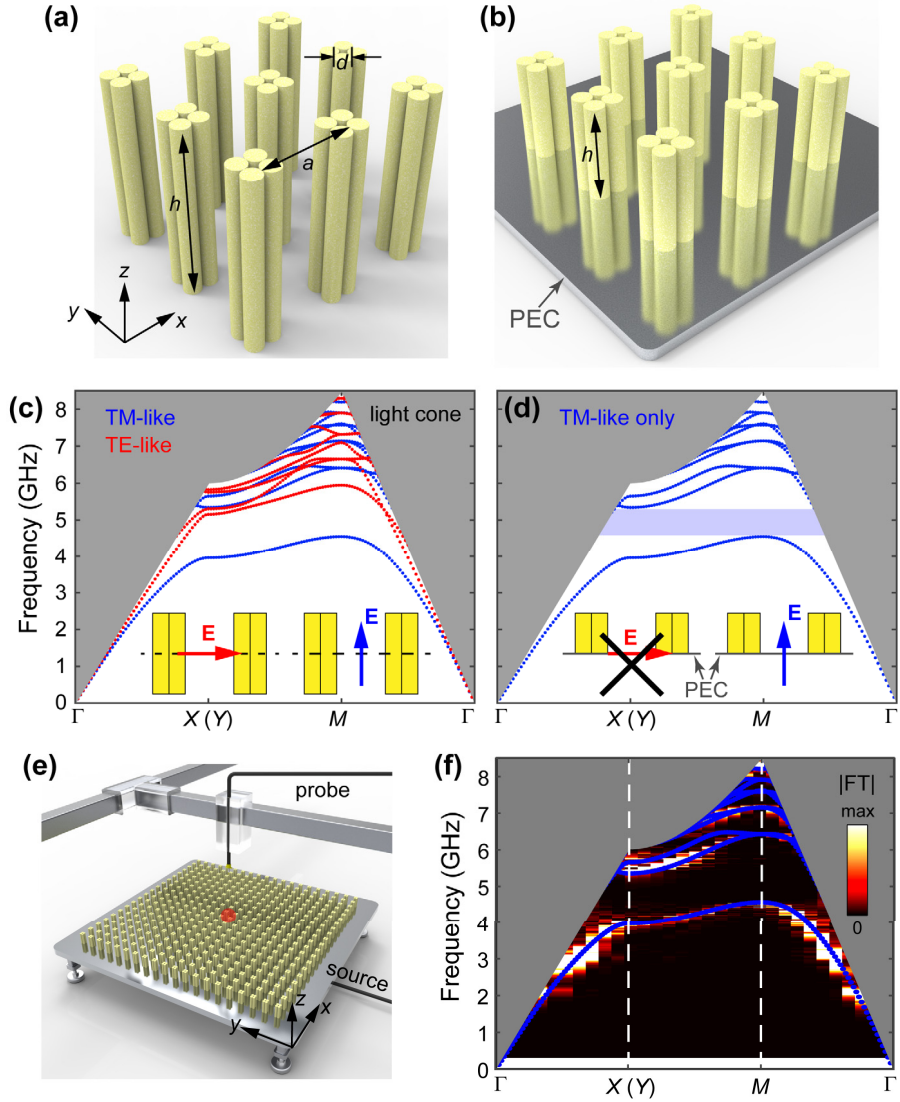

FIG. 1. (a) Schematics and (c) Bulk band structures of the free-standing PC slab. The in-plane lattice constant is *a*, the diameter and height of rods are *d* and *h*. Insets: Electric fields of TE-like and TM-like modes are in-plane and out-of-plane at the central plane. TM-like and TE-like bands are marked in blue and red, respectively. The light cone is shaded in grey. (b) Schematic and (d) Bulk band structures of the PC slab with rods standing on a PEC. Insets: Only TM-like modes exist as TE-like modes are filtered out by the PEC boundary condition. A full band gap is found (shaded in blue). (e) Experimental setup for the near-field scanning measurement. (f) Measured (bright color) and calculated (blue line) bulk band structures.

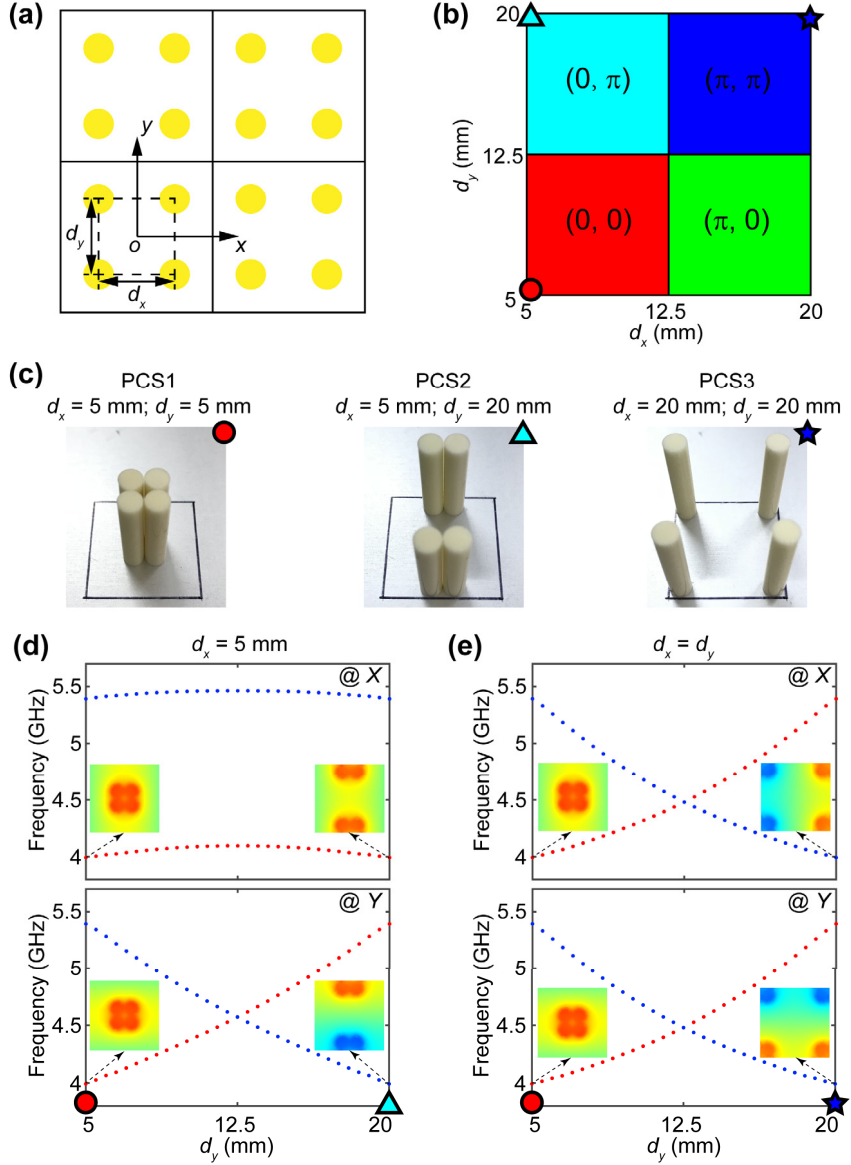

FIG. 2. (a) Schematic of the PC slab with four unit cells. The distances between two rods along the *x* and *y* directions are $d_x$ and $d_y$, respectively. (b) Classification of PC slabs based on the 2D Zak phase $\mathbf{Z} = (Z_x, Z_y)$. PC slabs with $d_x$ and $d_y$ ranging from 5 to 20 mm are classified into four regions. Three representative PC slabs are labelled by the red circle, cyan triangle and blue star. (c) Photos of three representative PC slabs with (left) $d_x$ = 5 mm, $d_y$ = 5 mm, (middle) $d_x$ = 5 mm, $d_y$ = 20 mm, (right) $d_x$ = 20 mm, $d_y$ = 20 mm. (d-e) Evolution of two lowest bulk states at the *X* and *Y* points as a function of $d_y$ whilst (d) keeping $d_x$ = 5 mm or (e) keeping $d_x = d_y$. Mirror symmetric (anti-symmetric) states are marked in red (blue). Insets: $E_z$ fields of the first lowest bulk states of each PC slab.

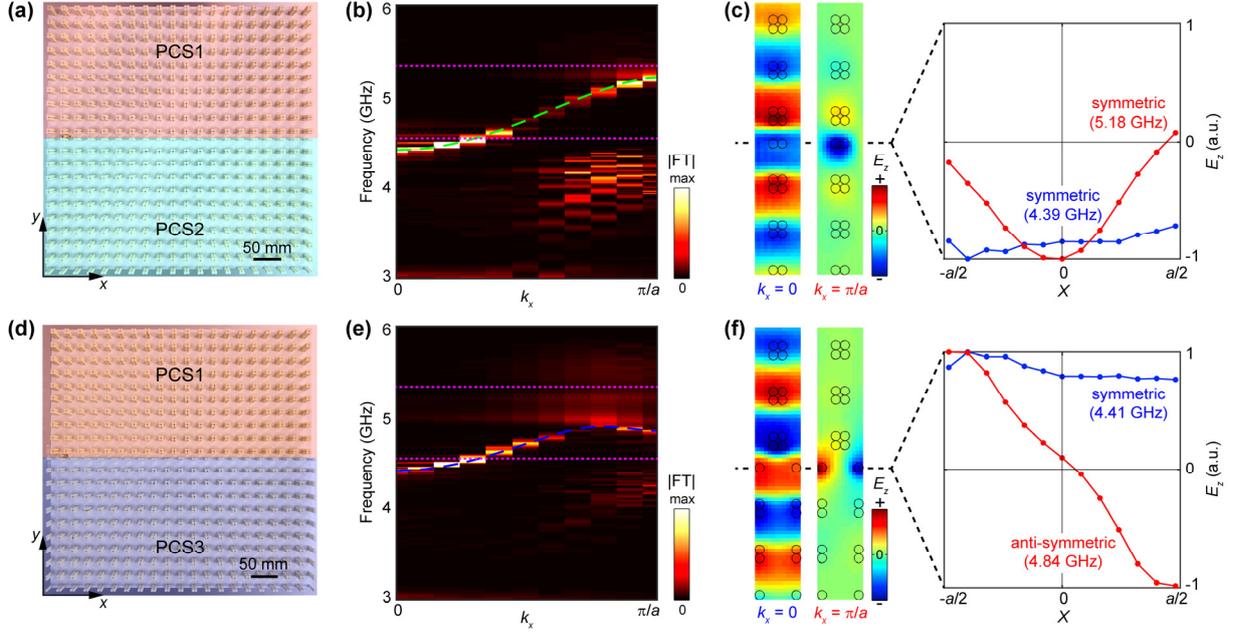

FIG. 3. (a) The photo of the boundary between PCS1 and PCS2. (b) The measured (bright color) and calculated (green line) gapped edge band dispersions. Two pink dash lines mark lower and upper frequencies of the bulk band gap. (c) The retrieved $E_z$ fields at the zone center ($k_x = 0$) and zone boundary ($k_x = \pi/a$). Both $E_z$ fields at the zone center and zone boundary are mirror symmetric. (d-f) Measured results for the boundary between PCS1 and PCS3. As shown in (f), edge state at zone center is mirror symmetric, while that at zone boundary is mirror anti-symmetric.

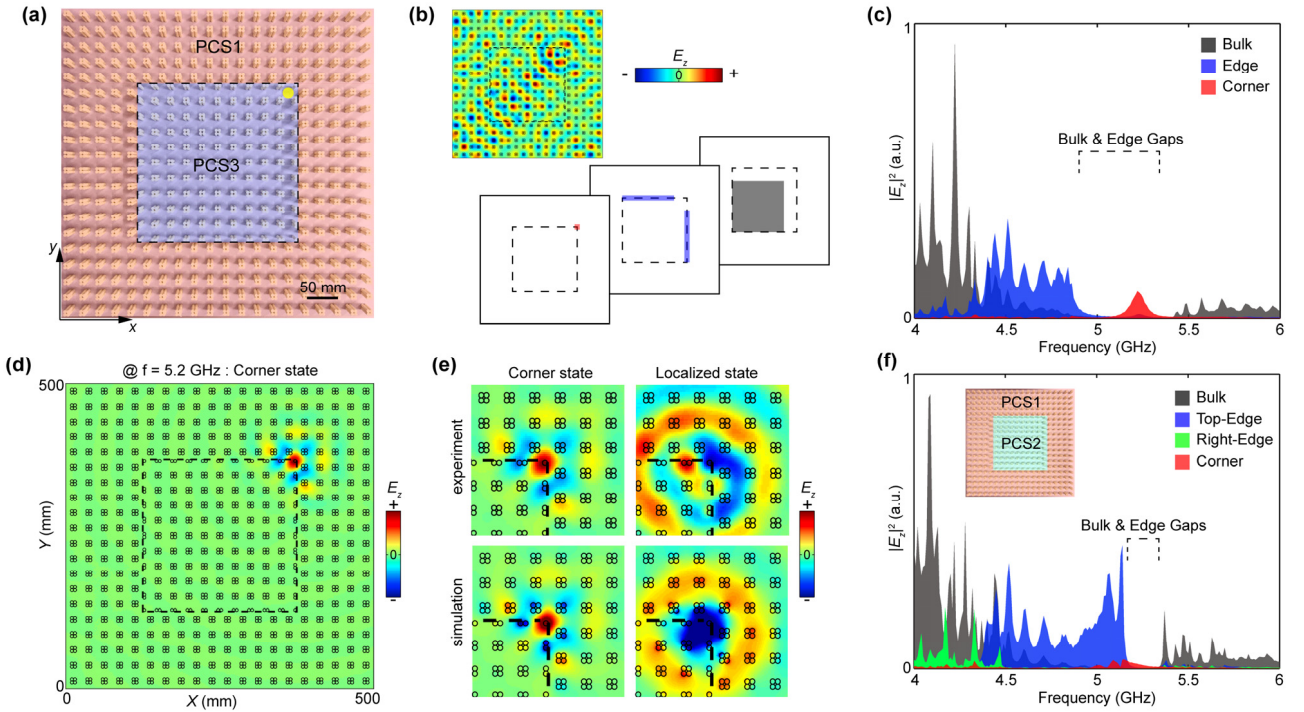

FIG. 4. (a) The photo of the experimental sample for measuring corner states. It consists of PCS3 at the center while PCS1 at the background, with the black dash square indicating boundaries. The yellow dot marks the position of the source antenna. (b) The response of the sample at one example frequency

of 4.01 GHz. Measured $|E_z|^2$ are multiplied by three binary filters to determine the bulk (grey), edge (blue) and corner (red) responses. (c) The resulting spectra of bulk, edge, and corner states, showing different resonant frequencies. (d) Measured $E_z$ fields of the corner state at 5.2 GHz. (e) Measured and simulated $E_z$ fields of corner (localized) state are shown in the left (right) panel. (f) The bulk, edge, and corner spectra of the sample without corner states. Inset: sample constructing by PCS1 and PCS2.